\documentclass[10pt,aps,prl,nofootinbib,superscriptaddress,twocolumn,preprintnumbers,balancelastpage]{revtex4-1}
\usepackage{amsmath,amssymb,amsthm,amsopn}
\usepackage{graphicx}
\usepackage[T1]{fontenc} 
\usepackage{tikz-feynman}
\usepackage{epsfig}
\usepackage{psfrag}
\usepackage[colorlinks=true,allcolors=blue!65]{hyperref}
\usepackage{multirow}
\usepackage{empheq}
\usepackage{slashed}
\usepackage{graphicx}
\usepackage{url}
\usepackage{subfigure}
\usepackage{textcomp}
\usepackage{bm}
\usepackage{dcolumn}
\usepackage{color,xcolor}
\usepackage{ulem}
\usepackage{cancel}
\usepackage{braket}
\usepackage[title]{appendix}
\usepackage{lipsum}
\usepackage{mathptmx}
\DeclareMathAlphabet{\mathcal}{OMS}{cmsy}{m}{n}

\def\comment#1{}

\def\beq{\begin{equation}}
\def\eeq{\end{equation}}
\def\bea{\begin{eqnarray}}
\def\eea{\end{eqnarray}}

\begin{document}

\title{Heavy QCD axion model in light of pulsar timing arrays}

\author{Moslem Ahmadvand}
\email[]{ahmadvand@ipm.ir}
\affiliation{School of Particles and Accelerators,  Institute Research \\ in Fundamental Sciences (IPM)
P. O. Box 19395-5531, Tehran, Iran}
\affiliation{School of Physics, Damghan University, Damghan 3671645667, Iran}
\author{Ligong Bian}
\email[]{lgbycl@cqu.edu.cn}
\affiliation{Department of Physics and Chongqing Key Laboratory for Strongly Coupled Physics,  \\ Chongqing University, Chongqing 401331, China}
\affiliation{Center for High Energy Physics, Peking University, Beijing 100871, China}
\author{ Soroush Shakeri}
\email[]{s.shakeri@iut.ac.ir}
\affiliation{Department of Physics, Isfahan University of Technology, Isfahan 84156-83111, Iran}
\affiliation{ICRANet, Piazza della Repubblica 10, I-65122 Pescara, Italy}




\date{\today}

\begin{abstract}
Recently, pulsar timing array experiments reported the observation of a stochastic gravitational wave background in the nanohertz range frequency band. We show that such a signal can be originated from a cosmological first-order phase transition (PT) within a well-motivated heavy (visible) QCD axion model. Considering the Peccei-Quinn symmetry breaking at the TeV scale in the scenario, we find a supercooled PT, in the parameter space of the model, prolonging the PT with the reheating temperature at the GeV scale.

\end{abstract}

\maketitle
\section{I. Introduction}
Gravitational wave (GW) experiments have provided new paths to explore the Universe and examine models of physics. Recently, pulsar timing array (PTA) experiments, including the North American Nanohertz Observatory for Gravitational Waves (NANOGrav) \cite{NANOGrav:2023gor}, and the European Pulsar Timing Array (EPTA) \cite{Antoniadis:2023ott} along with the Parkes Pulsar Timing Array (PPTA) \cite{Reardon:2023gzh} and Chinese Pulsar Timing Array \cite{Xu:2023wog} 
released their latest results observing significant evidence of a GW following the Hellings-Downs pattern \cite{1983ApJ...265L..39H} in the angular cross-correlation of pulsar timing residuals, supporting a stochastic gravitational background (SGWB).

Various sources can be the origin of these signals, for instance, supermassive black hole binary mergers, have been proposed as astrophysical candidates \cite{NANOGrav:2023hvm,Antoniadis:2023zhi,Rajagopal:1994zj,Ellis:2023dgf,Bi:2023tib}. However, such stochastic signals may come from a relic cosmic background \cite{Madge:2023cak} similar to the cosmic microwave background (CMB). In addition, cosmological sources such as first-order phase transition (FOPT) around the QCD scale \cite{Ellis:2023tsl,Ratzinger:2020koh,Addazi:2023jvg,Ghosh:2023aum,Rezapour:2020mvi,Rezapour:2022iqq,Neronov:2020qrl,Ahmadvand:2017tue,Ahmadvand:2017xrw,Athron:2023mer,Han:2023olf,Li:2023bxy}, cosmic strings \cite{Blasi:2020mfx,Ellis:2020ena,Buchmuller:2020lbh,Lazarides:2023ksx,Wang:2023len,Bian:2022tju}, domain walls \cite{Bai:2023cqj,Kitajima:2023cek,Blasi:2023sej,Li:2023yzq,Bian:2022qbh,Babichev:2023pbf,Sakharov:2021dim,Zhang:2023nrs}, primordial black holes \cite{Depta:2023qst,Gouttenoire:2023nzr,Franciolini:2023pbf,Guo:2023hyp,HosseiniMansoori:2023mqh} and inflation \cite{Firouzjahi:2023ahg,Vagnozzi:2020gtf,Ashoorioon:2022raz,Niu:2023bsr,Servant:2023mwt} may fit the recent data better \cite{NANOGrav:2023gor}. There have also been some recent discussions about the lake of strong preferences for any specific SGWB sources with the current data based on Bayesian analysis \cite{Bian:2023dnv}. However, future PTA datasets have the potential to distinguish between the cosmological and astrophysical scenarios, for example, by anisotropies, circular polarization or individual sources \cite{Ellis:2023oxs}.

An intriguing candidate that we propose in this paper is a supercooling phase transition (PT) which ends around the $\mathrm{GeV}$ scale; such a PT can naturally be accommodated in the context of heavy QCD axion models \cite{Rubakov:1997vp,DiLuzio:2020wdo}. A cosmic FOPT may be accomplished due to a spontaneous broken symmetry and the process is accompanied by the nucleation of bubbles separating true and false vacua. In a supercooled PT, the vacuum energy is dominated and the Universe remains in the false vacuum for a long period and supercools, increasing the PT duration \cite{Ellis:2019oqb,Brdar:2018num,Kobakhidze:2017mru,VonHarling:2019rgb,Athron:2023mer,Yang:2023qlf}.   

In this work, we consider a scenario within heavy axion models which are very appealing in that not only can the strong CP problem \cite{Peccei:1977hh} be addressed but also the models have richer phenomenology \cite{Berezhiani:2000gh,Dimopoulos:2016lvn} relative to the invisible QCD axion models \cite{Zhitnitsky:1980tq,Dine:1981rt,Kim:1979if,Shifman:1979if}. Moreover, the relation of the axion mass, $m_a$, and its decay constant, $f_a$ is modified so that larger $m_a$ and lower $f_a$ are allowed. These models are also motivated in connection with the Peccei-Quinn (PQ) quality problem \cite{Georgi:1981pu,Kamionkowski:1992mf,Holman:1992us} (see Appendix A).

We explore a supercooled PT with a U(1) PQ symmetry breaking at the TeV scale. Based on the nearly conformal dynamics of the PQ scalar field, we analytically find important parameters encoding the GW spectrum and show that the corresponding signals can be well fitted to the PTA data.

\section{II. PQ phase transition in the model}

The CP-violating $\bar{\theta}$ parameter in the strong interactions is experimentally bounded $\bar{\theta}\lesssim 10^{-10} $ \cite{Baker:2006ts} and the nature of this smallness, known as the strong CP problem, is a big puzzle in particle physics. One of the interesting solutions to this problem is based on a $U(1)$ global symmetry, $U(1)_{\mathrm{PQ}}$, first proposed by Peccei and Quinn \cite{Peccei:1977hh}. At energies higher than the electroweak (EW) scale, the symmetry is spontaneously broken and the axion as the pseudo-Nambu-Goldstone boson is generated. The interaction of the axion with gluons at the QCD scale, $ (a/f_a +\bar{\theta})G\widetilde{G} $, and as a result the vacuum expectation value of the axion can cancel the $\bar{\theta}$ term \cite{Peccei:2006as}. Considering the astrophysical constraints on the axion decay constant, $10^8\,\mathrm{GeV}\lesssim f_a \lesssim 10^{17}\,\mathrm{GeV}$ \cite{Raffelt:2006cw,Arvanitaki:2009fg,Shakeri:2022usk}, there are two classes of models, known as invisible axion models \cite{Zhitnitsky:1980tq,Dine:1981rt,Kim:1979if,Shifman:1979if}. However, as mentioned before, $f_a$ can be lowered, $f_a\sim (1-100)\,\mathrm{TeV}$ in the so-called visible axion models, still addressing the strong CP problem. In this sense, there are diverse scenarios, e.g. enlarging the QCD color gauge group $SU(3+\mathcal{N})$ \cite{Gherghetta:2016fhp} or assuming some hidden sector with a confining scale larger than the QCD cutoff $\Lambda_{\mathrm{QCD}}$ such that additional terms are supplemented to the axion potential and change the axion mass \cite{Berezhiani:2000gh,Fukuda:2015ana}. In the case of adding a hidden sector, one can consider a whole copy of the standard model (SM), with the $SM\times SM'$ gauge group, and particles of the two sectors can interact gravitationally or by some very feebly couplings. Furthermore, a $Z_2$ mirror symmetry between  particles of the two sectors can be imposed and soft terms breaking the symmetry can induce $\Lambda'_{\mathrm{QCD}}\gg \Lambda_{\mathrm{QCD}}$ \cite{Berezhiani:2000gh,Fukuda:2015ana}.

Here, we consider a DFSZ axion model \cite{Zhitnitsky:1980tq,Dine:1981rt}, containing a gauge-singlet PQ scalar field, $\Phi$, and two Higgs doublets under $SU(2)_L$, $H_u$ and $H_d$, which is supplemented with its hidden copy. Indeed, we consider a theory with the $SM\times SM'$ gauge group and each sector interacting with the PQ scalar (see Appendix B) and has the same PQ symmetry.\footnote{An analogous procedure can also be applied to a KSVZ axion model \cite{Kim:1979if,Shifman:1979if}.} Therefore, the axion couples to both QCD sectors
\begin{equation}
\frac{a}{f_a}\left(G\widetilde{G}+G'\widetilde{G}'\right),
\end{equation}
and its mass would be $m_a\sim \Lambda_{\mathrm{QCD}}'^2/f_a$ \cite{Berezhiani:2000gh,Fukuda:2015ana} with $f_a\sim \mathrm{TeV}$.
In this work, we do not go through axion interactions and focus on the UV theory and study the PQ PT associated with the PQ symmetry breaking.
We explore a supercooled PQ PT along the direction of PQ scalar dynamic, $\langle\Phi\rangle= v_{\phi}/\sqrt{2} $. We consider the case where mass parameters are small, $ \mu_d^2, \mu_u^2, \lambda_{\phi}f_a^2\ll f_a^2$ (see Appendix B) with the similar condition in the hidden sector, and main contributions to the potential arises from the one-loop Coleman-Weinberg (CW) quantum correction \cite{Coleman:1973jx} and the leading contribution from thermal corrections \cite{Quiros:1999jp}; thereby masses and the symmetry breaking scale are generated radiatively \cite{Gildener:1976ih}.

\begin{figure}[h]
	\centering
	\includegraphics[width=3.4in]{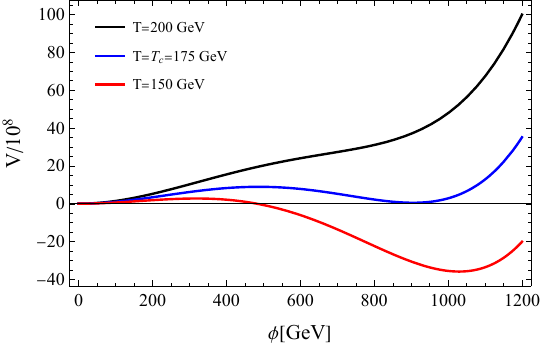}
	\caption{For three different values of temperature as labeled and representative values of $A=-0.0053$, $B=0038$, and $D=0.09$, the radiatively induced potential is illustrated. Above $T_c$, the origin is the only minimum. At $T_c$, two degenerate states appear and at lower temperatures the nonzero minimum is favored. The Universe remains in the false vacuum until the tunneling process is well developed at low temperatures due to $\lambda (T)$.}\label{fig1}
\end{figure}

At the zero-temperature limit, quantum corrections (see Appendix C) contribute to the potential that is approximately scale invariant as
\begin{equation}
V=\left(A+B\ln\frac{\phi^2}{f_a^2}\right)\phi^4\equiv \lambda(\phi)\phi^4. 
\end{equation}
where
\begin{equation}
	A=\frac{1}{64\pi^2}\left(\kappa_{u}^2\left(\ln(\kappa_{u})-\frac{3}{2}\right)+\kappa_u'^{2}\left(\ln(\kappa'_{u})-\frac{3}{2}\right)\right),
\end{equation}
\begin{equation}
	B=\frac{1}{64\pi^2}\left(\kappa_u^2+\kappa_u'^{2}\right).
\end{equation}
For the sake of simplicity, we only considered $\kappa_u$ and $\kappa_u'$ couplings which are determined at the vacuum and are then evolved to a scale $\phi$ through the beta function (see Appendix C). Thus, at the nonzero vacuum, $ v_{\phi}(T=0)\equiv f_a $, from $(dV/d\phi) |_{f_a}=0$ we find
\begin{equation}\label{min}
-4\lambda=\frac{d\lambda}{d\ln \phi},~~~~~\Delta V\equiv V(0)-V(f_a)=\frac{Bf_a^4}{2}=\frac{(\bar{\kappa}_u^2+\bar{\kappa}_u'^{2})f_a^4}{128\pi^2}.
\end{equation}
Considering thermal corrections (see Appendix C) and high temperatures, the origin point would be the minimal, and the potential would be
\begin{equation}\label{poten}
V=DT^2\phi^2+\lambda\phi^4+\cdots,  
\end{equation}
where 
\begin{equation}
D=\frac{1}{24}\left(\kappa_u+\kappa_u'\right).  
\end{equation}
As temperature goes down, negative values of $\lambda$ can change the potential slope and induce a bump so that at the critical temperature, $T_c$, two degenerate states, $\Delta V(T)= V(0, T)-V(v_{\phi}(T), T)=0$, indicating a FOPT, can be generated due to these values of $\lambda$, Fig.\ (\ref{fig1}). At temperatures below $T_c$, the vacuum $v_{\phi}(T)$ is the favorable one and at some temperature $T_n$ bubbles of the new phase are nucleated. In fact, bubble nucleation occurs when the bubble formation probability per unit Hubble space-time volume, which is proportional to $ \exp \left(-S_3(T)/T\right)$ where $S_3(T)$ is the bounce action quantifying the tunneling process, would be of the order of 1. As a result, we can find the nucleation temperature $T_n$ via the relation \cite{Linde:1981zj,Arnold:1991cv}
\begin{equation}\label{nuc}
	\frac{S_3(T_n)}{T_n}\sim 4\ln\Big(\frac{T_n}{H(T_n)}\Big), 
\end{equation}
where the Hubble parameter $H$ is given by
\begin{equation}
	H^2=\frac{1}{3 M_{\mathrm{Pl}}^2}\left(\xi T^4+\Delta V(T)\right),
\end{equation}
where the reduced Planck mass is $ M_{\mathrm{pl}}\simeq 2.43\times 10^{18}\,\mathrm{GeV}$, $\xi= \pi^2g_*/30$, and $g_*\simeq 107$ is the effective number of relativistic degrees of freedom, assuming that it does not change during the PT. At low temperatures, $T\ll T_c$, it is expected $v_{\phi}(T)=f_a$, and $\Delta V(T)$ is well approximated by $\Delta V$. Furthermore, in this limit, the bounce action can be well approximated by \cite{Witten:1980ez}
\begin{equation}
	\frac{S_3}{T} \simeq -18.9 \frac{\sqrt{2 D}}{\lambda(T)} \simeq -5.4 \frac{\sqrt{\kappa_u+\kappa_u'}}{\lambda(T)}   
\end{equation} 
and hence we can find the nucleation temperature as
\begin{equation}\label{nuc}
	4\ln\Big(\frac{\sqrt{3}M_{\mathrm{Pl}} T_n}{\sqrt{\xi T_n^4+\Delta V}}\Big)\simeq -5.4 \frac{\sqrt{\kappa_u+\kappa_u'}}{\lambda(T)}.   
\end{equation}
At the time when the transition completes, from the conservation of energy, one obtains $\rho_R(T_*)=\rho_R(T_p)+\Delta V $ \cite{Ellis:2019oqb}  where $\rho_R(T)=\xi T^4 $ is the radiation energy and $T_p$ is the percolation temperature. Assuming $T_p\simeq T_n$, the reheating temperature can be computed by
\begin{equation}
T_*^4=\frac{\Delta V }{\xi} +T_n^4. 
\end{equation}
For the case of fast reheating $H_*\equiv H(T_*)\simeq H(T_n)$. 
Another important quantity in characterizing the generated GWs is the inverse of PT duration calculated by the following relation
\begin{equation}\label{beta}
\frac{\beta}{H_*}=T_n\frac{d}{dT}\left(\frac{S_3(T)}{T}\right)\Bigg |_{T_n}\simeq -\frac{S_3(T_n)}{T_n}\frac{\beta_{\lambda}(T_n)}{\lambda(T_n)}, 
\end{equation}
where $(d\lambda(T)/d\ln T)|_{T_n}=\beta_{\lambda}(T_n)$. From Eq.\ (\ref{poten}) and $ V(\phi_0, T)=0$, for $T\ll T_c$, and $ \phi_0\sim T$ we obtain $\lambda(T_n)\sim A+B\ln (T_n/f_a) $. Therefore, expressing  $-\lambda(T_n)/\beta_{\lambda}(T_n)\sim \gamma\ln (f_a/T_n)$, we find
\begin{equation}\label{beta}
\frac{\beta}{H_*}\simeq \frac{4}{\gamma\ln (f_a/T_n)}\ln\Big(\frac{\sqrt{3}M_{\mathrm{Pl}} T_n}{\sqrt{\xi T_n^4+\Delta V}}\Big). 
\end{equation}
Expecting $\beta/H_*\sim \mathcal{O}(1-10)$ for the case of supercooling, we fix the parameter $\gamma$ with the data. 

The strength of the supercooled PT is also obtained by
\begin{equation}\label{al}
	\alpha =\frac{\Delta V}{\rho_R(T_n)}.
\end{equation}
In the next section, calculating the aforementioned quantities, we obtain the GW energy density spectrum.  

\section{III. SGWB signal}
In this section, based on the obtained PT quantities, we calculate the GW spectrum of the supercooled PT. In this sense, sources contributing to the GW production are bubble wall collisions and fluid motions. Relying on the recent numerical computation \cite{Lewicki:2022pdb}, the present-day GW energy density spectrum is obtained by 
\begin{equation}
	h^2\Omega _{\mathrm{GW}}(f)=1.67\times 10^{-5}\Big(\frac{H_*}{\beta}\Big)^{2}\Big(\frac{\kappa \alpha}{1+\alpha}\Big)^2 \Big(\frac{100}{g_*}\Big)^{\frac{1}{3}}S(f),
\end{equation}
where the spectral shape of the GW is obtained by
\begin{equation}\label{spec}
	S(f)=\frac{\bar{A}(a+b)^c}{\left[b\left(\frac{f}{f_{p,0}}\right)^{-\frac{a}{c}}+a\left(\frac{f}{f_{p,0}}\right)^{\frac{b}{c}}\right]^c},
\end{equation}
and $\kappa$ is the fraction of the vacuum energy converted to the kinetic
energy of bubble walls, $\kappa_{\mathrm{vac}}\simeq 1/(1+5/(\beta R_{\mathrm{eq}}))$, and fluid motions $\kappa_{\mathrm{f}}=1-\kappa_{\mathrm{vac}}$ \cite{Lewicki:2022pdb}. As for the spectral shape, Eq.\ (\ref{spec}), the spectrum's low-and high-frequency power-law tails are controlled by $a$ and $b$, its transition width from one power law to another is determined by $c$, and its amplitude depends on $\bar{A}$.

In our case with $\alpha\gg 1 $, bubbles collide in the vacuum and the bubble wall velocity can reach the speed of light; thus we consider $\kappa_{\mathrm{vac}}\simeq 1$ corresponding to $\beta R_{\mathrm{eq}}\gg 1$.
Therefore, according to Table I of Ref. \cite{Lewicki:2022pdb}, corresponding fit parameters of the spectral function appropriate for a global broken symmetry, $T_{rr}\propto R^{-2}$ (where $T_{rr} $ is the maximum of the stress-energy tensor and $R$ is the bubble radius), would be as $\bar{A}\simeq 5.93\times 10^{-2}$, $a\simeq 1.03$, $b\simeq 1.84$, and , $c\simeq 1.45$, whereas for the envelope approximation,  $\bar{A}\simeq 3.78\times 10^{-2}$, $a\simeq 3.08$, $b\simeq 0.98$, and , $c\simeq 1.91$.

The present redshifted peak frequency is given by
\begin{equation}
	f_{p,0}=16.5\times 10^{-6}[\mathrm{Hz}] \Big(\frac{f_p}{\beta}\Big)\Big(\frac{\beta}{H_*}\Big)\Big(\frac{T_*}{100~\mathrm{GeV}}\Big)\Big(\frac{g_*}{100}\Big)^{\frac{1}{6}},
\end{equation}
where for the case of $T_{rr}\propto R^{-2}$, $f_p/\beta \simeq 0.64/(2\pi)$ and for the envelope approximation $f_p/\beta \simeq 1.33/(2\pi)$ \cite{Lewicki:2022pdb}.

\begin{figure}[h]
	\centering
	\includegraphics[width=3.8in]{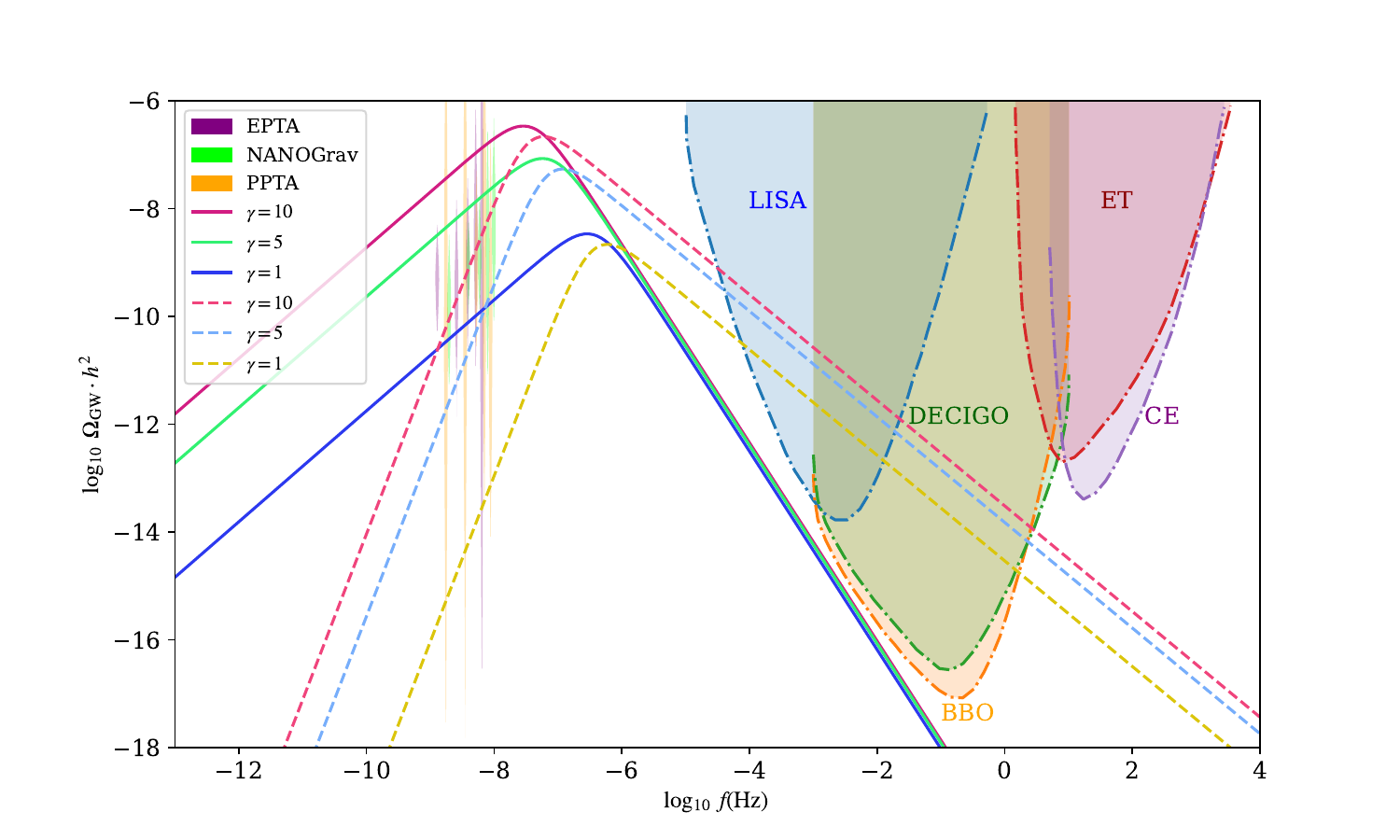}
	\caption{ GW spectra with the spectral shape for the case $T_{rr}\propto R^{-2}$ (solid line) and for envelope approximation (dashed line) are displayed. Three different values $\gamma=1, 5, 10$ correspond to the inverse duration of the PT $\beta/H_*=17, 3.4, 1.7$, respectively. The recent results of PTA experiments are also shown. The shaded regions represent  the sensitivity range of ongoing GW experiments:  LISA \cite{2017arXiv170200786A}, DECIGO \cite{2011CQGra..28i4011K}, BBO \cite{Crowder:2005nr}, Einstein Telescope (ET) \cite{Punturo:2010zz} and Cosmic Explorer (CE) \cite{Reitze:2019iox}.}\label{fig2}
\end{figure}

As can be seen from  Fig.\ \ref{fig2}, using datasets of NANOGrav 15 yr, EPTA release 2, and PPTA  data release 3 (DR3), for $f_a=1\,\mathrm{TeV}$, and some representative values of parameters, $\bar{\kappa}_u=\bar{\kappa}'_{u}=0.0001$, $\kappa_u=\kappa'_{u}=0.001$ and $\lambda(T)=-0.0015$, which correspond to $T_n=100\,\mathrm{MeV}$, $T_*=1\,\mathrm{GeV}$, $\alpha=8995$, and $\beta/H_*=17, 3.4, 1.7$ with the corresponding parameters $\gamma=1, 5, 10$, respectively, the GW signals can be very consistent with the PTA data from the bubble collision spectra for both the spectral shape of $T_{rr}\propto R^{-2}$ (solid line)  and for the envelope (dashed line). Remarkably, we see that the high-frequency range of the GW signals with the envelope approximation falls within the sensitivity range of future space-based GW experiments such as Laser Interferometer Space Antenna (LISA), Deci-hertz Interferometer Gravitational Wave Observatory (DECIGO), and Big Bang Observer (BBO) and can be probed by these detectors. 

\begin{figure}[h]
	\centering
	\includegraphics[width=3.5in]{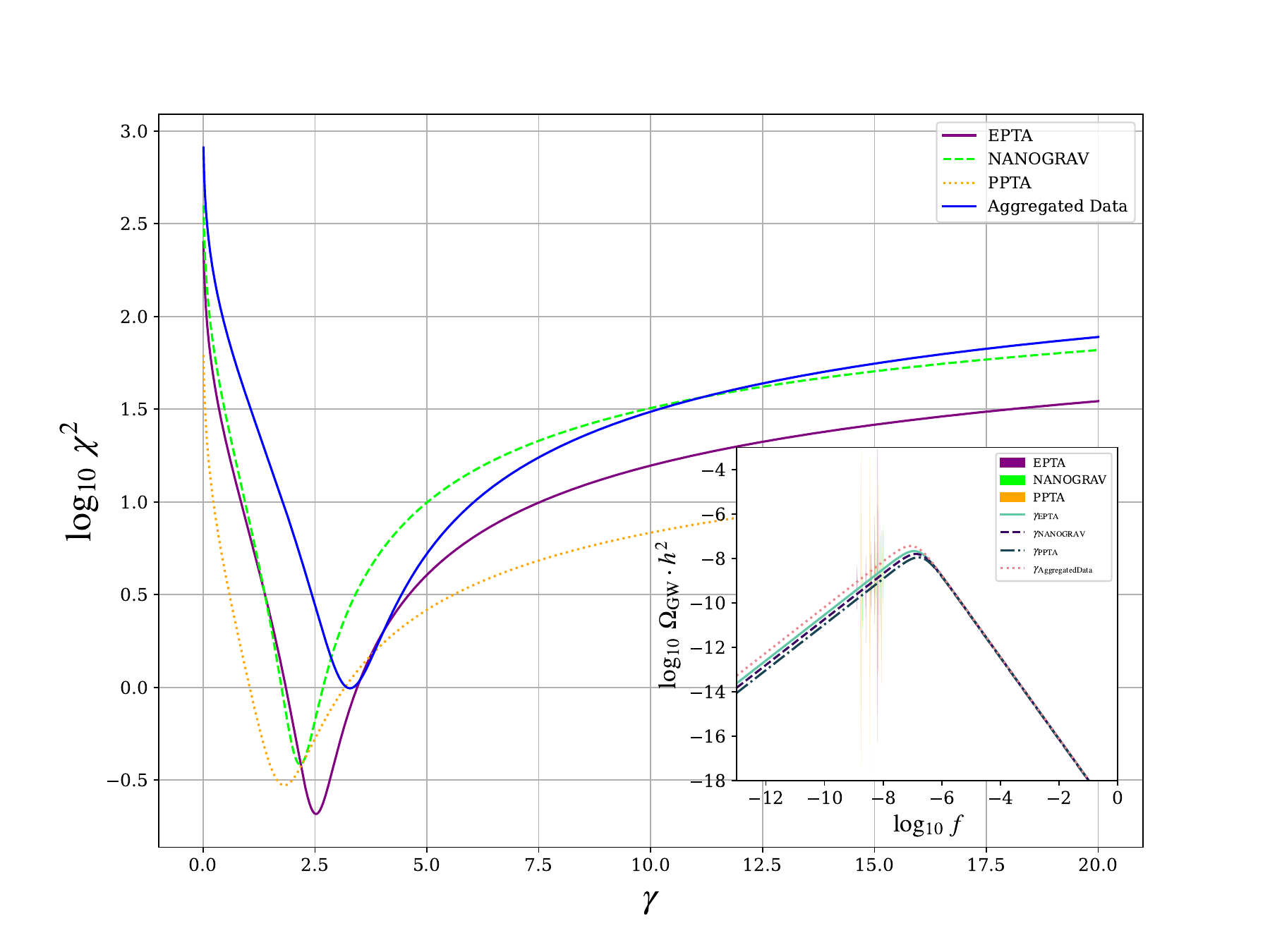}
	\caption{Based on the GWs with the spectral shape of $T_{rr}\propto R^{-2}$, we the find best-fit point in terms of $\gamma$ parameter, which is 2.52 for EPTA (solid purple), 2.17 for NANOGrav (dashed green), and 1.81 for PPTA (dotted orange). The best-fit value for the aggregated dataset is given by 3.28 (solid blue) corresponding to $\beta/H_*=5.1$. The insert shows GW spectra for the best-fit values of $\gamma$ in addition to observed signals. }\label{fig4}
\end{figure}

\begin{figure}[h]
	\centering
	\includegraphics[width=3.5in]{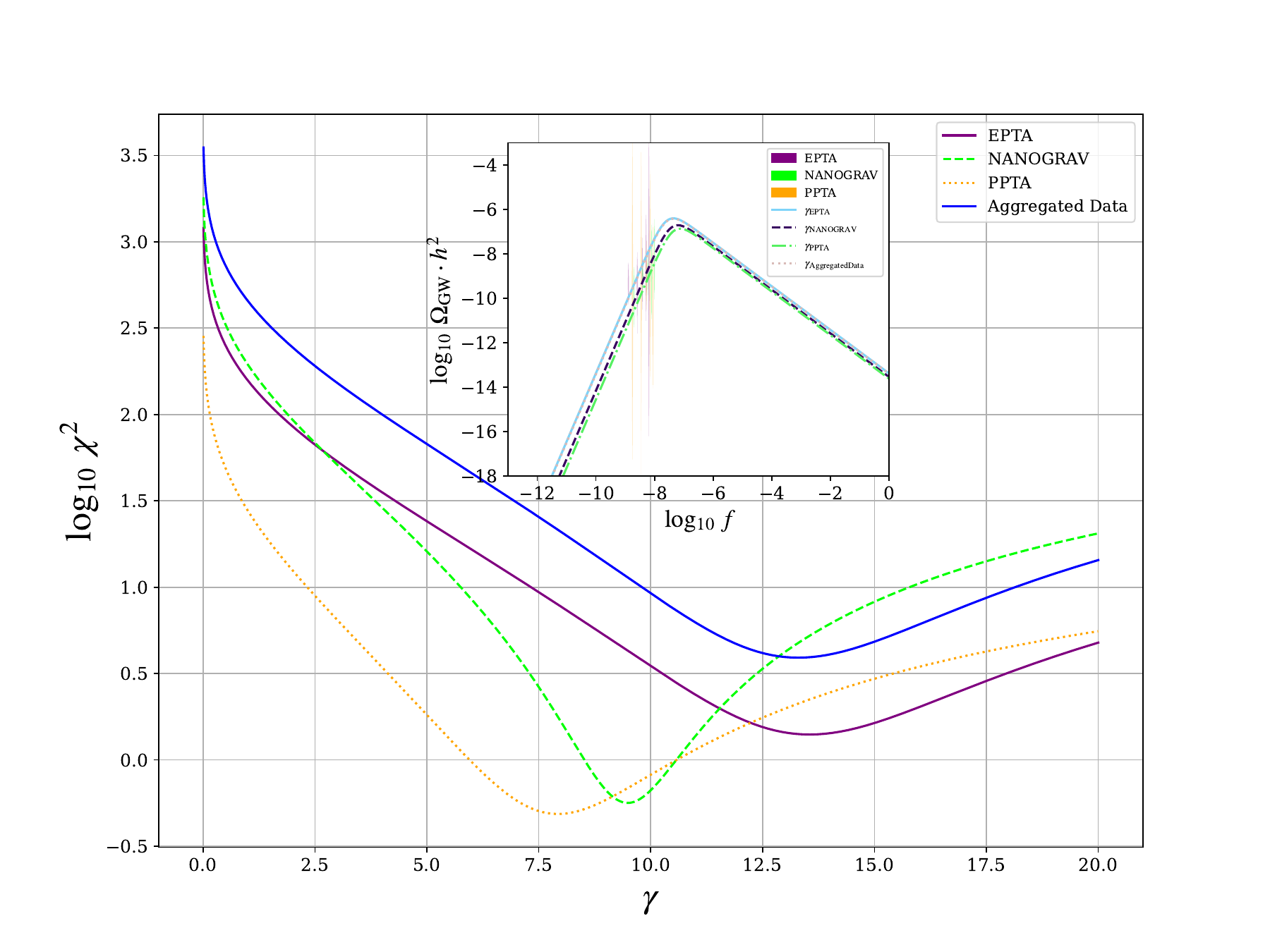}
	\caption{Based on the GWs with the envelope approximation.we find the best-fit point in terms of $\gamma$ parameter, which is 13.55 for EPTA (solid purple), 9.49 for NANOGrav (dashed green), and 7.93 for PPTA (dotted orange). The best-fit value for the aggregated dataset is given by 13.30 (solid blue) corresponding to $\beta/H_*=1.3$. The insert shows GW spectra for the best-fit values of $\gamma$ in addition to observed signals.}\label{fig5}
\end{figure}

Furthermore, using the PTA data, we perform a best-fit analysis over a range of $ \gamma$ parameter ($\beta/H_*$) through the $\chi^2$ test based on the relation 
\begin{equation}
\chi^{2}=\sum_{i=1}^{N} \frac{\left(\log_{10}\Omega_{\mathrm{th}}h^{2}-\log_{10}\Omega_{\mathrm{exp}}h^{2}\right)^{2}}{2\bar{S_{i}}^{2}},
\end{equation}
where $\Omega_{\mathrm{th}}h^2$ denotes the GW predicted by the model, $\Omega_{\mathrm{exp}}h^2$ represents the observed GW signal by PTA experiments, and $\bar{S_{i}}$ is  the deviation from the midpoint value of each data point in $\log_{10}\Omega_{\mathrm{exp}}h^2$ within the uncertainty range. As shown in Figs.\ \ref{fig4}, \ref{fig5} the best-fit point of the GW energy spectrum taking into account all datasets derived from the  $T_{rr}\propto R^{-2}$ case for the aggregated dataset is $\gamma=3.28$ ($\beta/H_*=5.1$) and for the envelope case, the best-fit point would be $\gamma=13.3$ ($\beta/H_*=1.3$).

\section{IV. Discussion}

Another PQ PT consequence of the model is that the symmetry breaking may induce a similar or different EW symmetry breaking scale in the ordinary and hidden sectors. For instance, as for  $v_{\mathrm{EW}}'$, this can be realized by these terms of the Lagrangian $\kappa'_{u} h_u^2 f_a^2+\lambda'_u h_u^4$, expressed in terms of the electrically neutral component of the fields, and hence $v_{\mathrm{EW}}'^2\simeq \kappa'_{u} f_a^2/\lambda'_u$. Setting $\kappa'_{u}=0.0001$ and $\lambda'_{u}=0.1$, we obtain $v_{\mathrm{EW}}'=1\,\mathrm{GeV}$. Even for these values of $v_{\mathrm{EW}}'$ its QCD scale could be $\Lambda'_{\mathrm{QCD}}\gtrsim v'_{\mathrm{EW}}$ \cite{Fukuda:2015ana}. Consequently, in this case, we obtain the axion mass $m_a\gtrsim 1\,\mathrm{MeV}$, while considering $ v_{\mathrm{EW}}\lesssim v'_{\mathrm{EW}}<f_a$ and assuming $\Lambda'_{\mathrm{QCD}}\lesssim v_{\mathrm{EW}} $, one finds $m_a \lesssim  10\,\mathrm{GeV}$.

Heavy axions may interact with visible and hidden (dark) photons 
\begin{equation}
\frac{a}{f_a}\left(F\widetilde{F}+F'\widetilde{F}'\right),
\end{equation}
so that the axion decay rate would be
\begin{equation}\label{drate}
\Gamma_{a \gamma \gamma}=\frac{g_{a \gamma}^2 m_a^3}{64 \pi}, \quad \Gamma_{a \gamma^{\prime} \gamma^{\prime}}=\frac{g_{a \gamma}^{\prime 2} m_a^3}{64 \pi}
\end{equation}
where

\begin{equation}
	g_{a\gamma}\simeq \frac{\bar{\alpha}}{2\pi}\frac{E}{N f_a}\frac{2z}{1+z},~~~~~~g_{a \gamma}^{\prime}\simeq \frac{\bar{\alpha'}}{2\pi}\frac{E}{N f_a}\frac{2z'}{1+z'},
\end{equation}
$\bar{\alpha}=1/137$ is the fine structure constant; $z=m_u/m_d$; and $E$ and $N$ are the electromagnetic and QCD anomaly coefficients, respectively, \cite{DiLuzio:2020wdo}. The heavy QCD axions can also couple to SM fermions
\begin{equation}
\frac{c_f}{2f_a}\partial_{\mu} a \bar{f}\gamma^{\mu}\gamma_5 f+\frac{c'_f}{2f_a}\partial_{\mu} a \bar{f'}\gamma^{\mu}\gamma_5 f'.
\end{equation}
Decaying the heavy axions to these photons contributes to the effective number of relativistic species which is strongly constrained by CMB measurements, $N_{\mathrm{eff}}=2.96^{+0.34}_{-0.33}$ \cite{Planck:2018vyg}, putting the most stringent constraints on the heavy QCD axion mass. According to the analysis of Ref.\ \cite{Dunsky:2022uoq}, for $E/N=8/3$ which may be the case of a DFSZ model  \cite{Ahmadvand:2021vxs}, $m_a<100\,\mathrm{MeV}$ would be excluded. However, lower masses, $1\,\mathrm{MeV}<m_a<100\,\mathrm{MeV}$ where the dark photon contribution to $N_{\mathrm{eff}}$ compensates neutrino dilution from axion decays to photons, are allowed for $E/N=1/3$.

We assume a baryon asymmetric dark sector. Thus, considering $ v_{\mathrm{EW}}\lesssim v'_{\mathrm{EW}}<f_a$, the mass range and relic abundance of stable particles in the hidden sector are approximately the same as those of the ordinary sector. As a result, the relic abundance of these particles cannot give rise to cosmological problems. Moreover, we assume that in the hidden sector there is no seesaw mechanism, which allows the Dirac neutrino mass in this sector as \cite{Fukuda:2015ana}
\begin{equation}
m_{v'}\sim \frac{v'_{\mathrm{EW}}}{v_{\mathrm{EW}}}\sqrt{M_R m_v},
\end{equation}
where $M_R$ is the right-handed neutrino mass in the ordinary sector. For $M_R\gtrsim 10^7\,\mathrm{GeV}$ and $m_a\gtrsim 1\,\mathrm{MeV}$ ($\Lambda'_{\mathrm{QCD}}\gtrsim 1\,\mathrm{GeV} $), the hidden neutrinos decay to the hidden electron and pion, $v' \rightarrow e'+\pi'$, and thereby these neutrinos do not contribute to $N_{eff}$.

Concerning astrophysical bounds, the model parameters can be constrained from SN 1987A \cite{Raffelt:1990yz}, with the supernova core cooling rate due to axion emission. According to Eq.\ (\ref{drate}), 
one can obtain the axion mean free path as
\begin{equation}
\lambda_{mfp}= \frac{v E_{a} }{m_{a}\Gamma_{a\gamma\gamma}}\approx  10 \ \text{m} \left( \frac{f_{a}}{\text{TeV}}\right)^{2} \left(\frac{E_{a}}{200\ \text{MeV}}\right) \left(\frac{100\text{MeV}}{m_{a}} \right)^4
\end{equation}
where $E_{a}$ stands for typical energies of the axions. Hence, for $E_{a}\gtrsim 200\,\mathrm{MeV}$, $f_a=1\,\mathrm{TeV}$ and $m_a\gtrsim 100\,\mathrm{MeV}$, we find that $\lambda_{mfp}$ is much smaller than the approximate radius of the supernova core $R\sim 10\,\mathrm{km}$. Thus, because of  the scattering and reabsorption, the axions are trapped within the supernova, and cannot take  away any energy \cite{Essig:2010gu,Jaeckel:2017tud}. 

In addition, the emission of dark photons can lead to a very fast core cooling rate of the supernova and the associated energy loss rate, which is bounded $L_{\gamma '}<10^{52}\,\mathrm{erg}/\mathrm{s}$, and can be obtained as $L_{\gamma '}\simeq \Gamma_{a \gamma^{\prime} \gamma^{\prime}} m_a(4\pi R^3/3)(1.2T^3/\pi^2)$ \cite{Berezhiani:2000gh}. Thus, taking into account a core temperature $T\simeq 10\,\mathrm{MeV}$, $f_a=1\,\mathrm{TeV}$, $m_a= 1\,\mathrm{MeV}$, and $z'=z\simeq 0.5$, the bound can be satisfied, while for $m_a\gtrsim 100\,\mathrm{MeV}$, it may require $z'\lesssim 10^{-4}$.

Considering the axion-electron interactions, the astrophysical bounds on helium burning lifetime of the horizontal branch stars imply $g_{ae}\sim m_e/f_a<(2.5-6)\times 10^{-13}$ \cite{Berezhiani:2000gh,DiLuzio:2020wdo}, and hence restrict $f_a $ to its invisible values. However, this bound does not apply to the heavy axions with $m_a>300\,\mathrm{keV}$. In fact, in the stellar cores whose typical temperatures reach $10\,\mathrm{keV}$, the axion production rate is suppressed by the exponential factor $\exp(-m_a/T)$. Therefore, astrophysical bounds allow heavy axions with masses above  $100\,\mathrm{MeV}$ for $f_a=1\,\mathrm{TeV}$. 

The viable parameter space of the model can also be probed via other experiments such as proton beam dump \cite{CHARM:1985anb,Aloni:2018vki}, kaon decays \cite{Alves:2017avw,KOTO:2018dsc}, and colliders \cite{Knapen:2016moh}. For $m_a\gtrsim 3m_{\pi} $, the heavy axions may be produced via $B\rightarrow K\,a$, where $B$ and $K$ denote $B$ mesons and kaons, respectively, and then mainly decay into three pions. For $m_a\lesssim 3m_{\pi} $, the dominant decay channels are the axion decay into two photons, electrons, and muons, for $m_a>2m_e$ and $m_a>2m_{\mu}$, respectively. In this sense, $m_a\lesssim 400\,\mathrm{MeV}$ for $f_a$ at the TeV scale is excluded by experiments such as kaon decays and the proton beam dump, while for $m_a\gtrsim 400\,\mathrm{MeV}$ the search is challenging due to the dominance of the hadronic decay mode \cite{Bertholet:2021hjl}.

It is noteworthy that the heavy axion cannot be a dark matter candidate in this context. However, some stable particles in the hidden sector may contribute to the dark matter relic abundance.\footnote{Moreover, according to Eq.\ (\ref{pot}), the interaction of the Higgs boson with the heavy axions can contribute to its invisible decays \cite{ATLAS:2022yvh,CMS:2022qva}, and since the heavy axions are unstable, they can decay to stable particles in the hidden sector}. We leave detailed calculations in connection with axion interactions and possible dark matter candidates in the presented scenario to a future work.

\section{V. Conclusion}
With regard to the recently detected GW signals by PTA experiments, we have proposed a heavy QCD axion scenario, with a PQ symmetry breaking scale around the TeV scale, based on a (DFSZ) axion model which is supplemented with its hidden copy. We investigated a supercooled PQ FOPT which is derived by CW quantum corrections and analytically found important PT quantities, including the reheating temperature, the inverse duration of the PT, and the strength of the PT. We have shown that within the parameter space of the model the generated GWs from the bubble wall collisions can be consistently fitted with the recently observed PTA data. Furthermore, it is shown that the high-frequency range of such GW signals can be probed by future space-based GW detectors. 

In addition, Heavy axion models are well-motivated scenarios allowing the $(m_a, f_a)$ relation to be relaxed, still addressing the strong CP problem. It is shown that considering the EW symmetry breaking scale in the ordinary and hidden sectors induced after the PQ symmetry breaking, a range of axion masses, $1\,\mathrm{MeV}\lesssim m_a \lesssim 10\,\mathrm{GeV}$ can be obtained. Nevertheless, viable regions of the mass space are limited via CMB observation due to the dark photon contribution to $N_{\mathrm{eff}}$, and also other experimental setups such as the proton beam dump.

As a result, interpreting the recent GW data, the model provides a setup which can be probed by future GW detectors, CMB telescopes and collider experiments.

\section{ACKNOWLEDGMENTS}

SS thanks Fazlollah Hajkarim and Seyed Mohammad Mahdi Sanagostar for helpful discussions about data analysis. This work is supported in part by the National Key Research and Development Program of China under Grant No. 2021YFC2203004, and the National Natural Science Foundation of China (NSFC) under Grants No. 12075041 and No. 12147102. 

\section{Appendix A: The PQ quality problem}
Invisible axion models with $ f_a\gtrsim 10^8\,\mathrm{GeV}$ suffer from the axion quality problem. Indeed, it is expected that the PQ symmetry is explicitly broken by higher-dimensional Planck-suppressed operators which can spoil the axion solution. In our model, the PQ symmetry may be explicitly broken by the following five-dimensional operator
\begin{equation}
	\frac{\lambda_a}{5!M_P}\phi_a^5
\end{equation}
where $M_P$ denotes the Planck mass and $\phi_a$ is the radial component of the PQ scalar. Such an explicit PQ symmetry breaking term induces a mass term in the axion field, $m_*^2\sim 10^{-2}\lambda_a (f_a^3/M_P) $, and moves the axion potential minimum away from the CP conserving minimum, $\Delta\bar{\theta}\sim m_*^2/m_a^2$ \cite{DiLuzio:2020wdo}. Therefore, assuming $\lambda_a=\mathcal{O}(1)$, for $f_a=1\,\mathrm{TeV}$ and $m_a\gtrsim 100\,\mathrm{MeV}$, the model satisfies the current bound $\Delta\bar{\theta}\lesssim 10^{-10}$.

\section{Appendix B: The scalar potential}\label{ap}
The tree-level scalar potential with the PQ symmetry is given by \cite{Ahmadvand:2021vxs}
\begin{align} \label{pot}	
	&V(\Phi, H_u, H_d) = \lambda_{\phi} \left(|\Phi|^{2} -f^{2}\right)^{2} + \left|H_{d}\right|^{2} \left(\kappa_{d}\, |\Phi|^{2} -\mu_{d}^{2}\right)\nonumber\\ &+ \left|H_{u}\right|^{2} \left(\kappa_{u}\, |\Phi|^{2} - \mu_{u}^{2}\right) -\left(\kappa_{ud}\, \Phi H_{u}^{\dagger}\, H_{d} + \text{H.c.}\right)\nonumber\\
	& + \lambda_{d} \left|H_{d}\right|^{4} + \lambda_{u} \left|H_{u}\right|^{4} + \lambda_{ud} \left(\left|H_{u}\right|^{2}\left|H_{d}\right|^2-\left|H_{u}^{\dagger} H_{d}\right|^{2} \right).
\end{align}
We also assume an analogous potential with $H'_{d}$ and $H'_{u}$ from the hidden sector. 

\section{Appendix C: Quantum and thermal corrections}\label{app}

The one-loop Coleman-Weinberg quantum correction is given by 
\begin{equation}
	\begin{aligned}
		V_{\mathrm{CW}}\left(\phi\right)&=\sum_{i}(-1)^{F_{b/f}} g_{i} \frac{m_{i}^{4}\left(\phi\right)}{64 \pi^{2}}\left[\ln \left(\frac{m_{i}^{2}\left(\phi\right)}{\Lambda^{2}}\right)-c_{i}\right]
	\end{aligned}
\end{equation}
where $ F_{b/f}=1(0)$ for fermions (bosons), $ g_i$ is the number of degrees of freedom for a given field, and $c_i=3/2(5/2)$ for scalars and fermions (vectors).

For our model, considering only $\kappa_{u}$ and $\kappa'_{u}$ couplings, $m^2(\phi)=\kappa_{u}\phi^2$ (and the same with $\kappa'_{u}$), one can write

\begin{equation}
	V_{\mathrm{CW}}=\left(A+B\ln\frac{\phi^2}{\Lambda^2}\right)\phi^4\equiv \lambda(\phi)\phi^4,
\end{equation}
where
\begin{equation}
	A=\frac{1}{64\pi^2}\left(\kappa_{u}^2\left(\ln(\kappa_{u})-\frac{3}{2}\right)+\kappa_u'^{2}\left(\ln(\kappa'_{u})-\frac{3}{2}\right)\right),
\end{equation}
\begin{equation}
B=\frac{1}{64\pi^2}\left(\kappa_u^2+\kappa_u'^{2}\right).
\end{equation}

The running couplings at the scale $\phi$ are also obtained by the leading-order beta function
\begin{equation}
\frac{d\kappa_u}{d\ln\phi}\simeq \frac{\kappa_u^2}{16\pi^2}+\frac{\kappa_u \kappa_u^{\prime}}{8\pi^2}
\end{equation}
\begin{equation}
\frac{d\kappa_u^{\prime}}{d\ln\phi}\simeq \frac{\kappa_u'^{2}}{16\pi^2}+\frac{\kappa_u \kappa_u^{\prime}}{8\pi^2}
\end{equation}
The one-loop thermal correction is expressed as
\begin{equation}
	V_{T}\left(\phi, T\right)=\sum_{i}(-1)^{F_{b/f}} g_{i} \frac{T^{4}}{2 \pi^{2}} J_{b / f}\left[\frac{m_{i}^{2}\left(\phi\right)}{T^{2}}\right]
\end{equation}
where 
\begin{equation}
	J_{b/f}\left(y^{2}\right)=\int_{0}^{\infty} d x\, x^{2} \ln \left[1 \mp e^{-\sqrt{x^{2}+y^{2}}}\right]
\end{equation}
and these thermal functions in the high-temperature limit are given by
\begin{align}\label{therm}
	J_{b}\left(\frac{m^{2}}{T^{2}}\right)&=-\frac{\pi^{4}}{45}+\frac{\pi^{2}}{12}\left(\frac{m}{T}\right)^{2}-\frac{\pi}{6}\left(\frac{m^{2}}{T^{2}}\right)^{3 / 2}\nonumber\\&-\frac{1}{32}\left(\frac{m}{T}\right)^{4} \ln \left(\frac{m^{2}}{a_{b} T^{2}}\right)+\cdots
\end{align}
\begin{equation}
	J_{f}\left(\frac{m^{2}}{T^{2}}\right)=\frac{7 \pi^{4}}{360}-\frac{\pi^{2}}{24}\left(\frac{m}{T}\right)^{2}-\frac{1}{32}\left(\frac{m}{T}\right)^{4} \ln \left(\frac{m^{2}}{a_{f} T^{2}}\right)+\cdots
\end{equation}
where $ \ln(a_b)=5.4076 $ and $ \ln(a_f)=2.6351 $. Thus, in the high-temperature limit one has
\begin{equation}
	V_{T}\left(\phi, T\right)=\frac{1}{24}\left(\kappa_u+\kappa_u'\right)T^2\phi^2+\cdots.
\end{equation}

\bibliography{references.bib}



\end{document}